# Origin of high hardness and optoelectronic and thermo-physical properties of boron-rich compounds B₆X (X = S, Se): a comprehensive study via DFT approach

M. M. Hossain[1*], M. A. Ali[1], M. M. Uddin[1], A. K. M. A. Islam[2,3], S. H. Naqib[2**]

[1]Department of Physics, Chittagong University of Engineering and Technology (CUET), Chattogram-4349, Bangladesh
[2]Department of Physics, University of Rajshahi, Rajshahi 6205, Bangladesh
[3]Department of Electrical and Electronic Engineering, International Islamic University Chittagong, Kumira, Chittagong, 4318, Bangladesh

Corresponding authors; *email: mukter_phy @cuet.ac.bd;
**email: salehnaqib@yahoo.com

**Abstract**

In the present study, the structural and hitherto uninvestigated mechanical (elastic stiffness constants, machinability index, Cauchy pressure, anisotropy indices, brittleness/ductility, Poisson's ratio), electronic, optical, and thermodynamic properties of novel boron-rich compounds $B_6X$ (X = S, Se) have been explored using density functional theory. The estimated structural lattice parameters were consistent with the prior report. The mechanical and dynamical stability of these compounds have been established theoretically. The materials are brittle in nature and elastically anisotropic. The value of fracture toughness, $K_{IC}$ for the $B_6S$ and $B_6Se$ are ~ 2.07 MPam$^{0.5}$, evaluating the resistance to limit the crack propagation inside the materials. Both $B_6S$ and $B_6Se$ compounds possess high hardness values in the range 31-35 GPa, and have the potential to be prominent members of the class of hard compounds. Strong covalent bonding and sharp peak at low energy below the Fermi level confirmed by partial density of states (PDOS) resulted in the high hardness. The profile of band structure, as well as DOS, assesses the indirect semiconducting nature of the titled compounds. The comparatively high value of Debye temperature ($\Theta_D$), minimum thermal conductivity ($K_{min}$), lattice thermal conductivity ($k_{ph}$), low thermal expansion coefficient, and low density suggest that both boron-rich chalcogenides might be used as thermal management materials. Large absorption capacities in the mid ultraviolet region (3.2-15 eV) of the studied materials and low reflectivity (~16 %) are significantly noted. Such favorable features give promise to the compounds under investigation to be used in UV surface-disinfection devices as well as medical sterilizer equipment applications. Excellent correlations are found among all the studied physical properties of these compounds.

## 1. Introduction

At present, materials with extraordinary physical and chemical properties are crying need in the development of modern industry to meet emergent technological demands. Unusual crystal structures sometimes reveal outstanding properties which are crucially important in advanced engineering and smart functional materials science. Materials like boron-rich systems belong to $B_{12}$ closed cage clusters (a three-dimensional rigid network) structure [1,2], which are quite uncommon. The boron and its compounds with $B_{12}$-icosahedral units have both fundamental and technological interest due to its exceptional properties such as high hardness with good mechanical strength, low density, very high melting point, and superior thermal and chemical stabilities. All these open a window for high-temperature and pressure device applications [3,4]. Specifically, many heavy-duty tools exposed to abrasive environment and high strain such as turbine and rotor blades, sports equipment, novel biological and chemical sensors, materials for armor are used extensively with boron-rich compounds [4,5]. Very recently in 2020 [6], boron-rich chalcogenide compounds, $B_6X$ (X = S, Se) with $B_{12}$ *closo*-clusters structure, have been synthesized at 6.1 GPa and 2700 K in a toroid shaped high-pressure apparatus as well as was predicted via calculating the formation enthalpy and phonon studies using Vienna *ab initio* Simulation Package (VASP). The study reported that the compounds have orthorhombic crystal structure with space group *Pmna*. Mechanical properties (polycrystalline elastic moduli and hardness) have been predicted as a member of family of hard phases with a hardness value of ~31 GPa [6]. Some boron sulfide compounds such as r-BS, $B_2S_3$ (II) and $B_2S_3$ (III) with different crystal symmetry groups have also been synthesized at high pressure and temperature conditions [3,4,7]. Zhang et al., [8] have studied the phase stability, mechanical and electronic properties of the Si–B system and other related phases: α-$SiB_3$($SiB_3$,$SiB_4$), β-$SiB_3$, $SiB_6$, $SiB_{36}$ using first-principles method. They have also reported that the bulk modulus and shear modulus are in the ranges 120 -180 GPa and 55-154 GPa, respectively. In an another work, the structural and thermoelectric properties of boron-rich solids, $B_6S_{1-x}$ (0.37 < x ≤ 0.40) in rhombohedral structure with $B_{12}$-icosahedral framework have been reported with semiconducting behavior and maximum Seebeck coefficient of 220 $\mu VK^{-1}$ [9]. Some other boron rich hexacarbide and silicide have also been known. Recently, a metastable structure of boron rich carbide $B_6C$ has been reported that exhibits superhard and superconducting properties [10]. Yuan et al., [11] have predicted a novel metallic silicon hexaboride $B_6Si$. The compound $B_6Si$ with space group *Cmca* has three dimensional structures of $B_{12}$ icosahedrons and Si atoms.

However, it is instructive to note that existing studies on the titled boron-rich compounds are limited only to structural, Raman spectroscopy and mechanical (elastic moduli and hardness) properties. There is significant lack of understanding of some other decisive properties concerning electronic band structure, optical, thermal and mechanical (elastic stiffness constant, fracture toughness, machinability index, Cauchy pressure, anisotropic behavior, brittleness/ductility, Poisson's ratio) features which must be attained to unlock the full potential of $B_6X$ (X = S, Se) for possible applications. First of all, knowledge on the electronic (band structure, electronic energy density of states, charge density distribution, Mulliken population analysis, electron localization function) and optical (absorption coefficient, reflectivity, photoconductivity, refractive indices, loss function and dielectric functions) properties are very crucial to design a photovoltaic and optoelectronic device applications. The knowledge of mechanical anisotropy is also of keen interest due to its close link to some important mechanical behaviors like crack formation and its subsequent propagation, instability of crystal structure, phase transformation, and growth of plastic deformations which sometimes restrict fruitful applications of materials. Again, prior knowledge on thermodynamic properties (Debye temperature, temperature dependent lattice thermal conductivity, thermal expansion coefficient, specific heats) and their correlations with mechanical properties could open a new window regarding potential device applications point of view under extreme conditions. Exploration of these uninvestigated properties constitutes the prime motivation of this work.

Therefore, a comprehensive study on electronic, optical, mechanical (unexplored part) and thermal properties of newly synthesized boron-rich chalcogenide compounds, $B_6X$ (X = S, Se) have been performed meticulously with the aid of density functional theory approach for the first time.

## 2. Computational methodology

Theoretical computations on the synthesized boron-rich chalcogenide compounds, $B_6X$ (X = S, Se) have been performed by first-principles method in the frame-work of state-of-the-art density functional theory [12,13], which is implemented in the CASTEP module [14,15]. The crystal structures are optimized using Broyden Fletcher Goldfarb Shanno (BFGS) optimization method with ultra-soft pseudopotential [16]. The computational parameters are as follows: plane wave basis set energy cut-off of 550 eV; Monkhorst–Pack [17] *k*-point mesh of size 4 × 5 × 3; convergence thresholds for the total energy, $5\times10^{-6}$ eV/atom; maximum force, 0.01 eV/Å; maximum stress, 0.02 GPa; maximum atomic displacement, $5\times10^{-4}$ Å. The ground state energy of the titled compounds was obtained using the Kohn-Sham (KS) equation. The solution KS equation was justified with electron exchange-correlations energy, where the functional form is of Perdew-Burke-Ernzerhof (PBE) type within the generalized gradient approximation (GGA) [13,18,19]. To make the computations comprehensive, the band structures of both the compounds were also calculated with local density approximation (LDA) and hybrid exchange-correlations functional HSE06 in addition to the GGA to obtain the fundamental band gap as accurately as possible. Other properties such as mechanical, optical, Mulliken population analysis and electron density difference are calculated using the GGA-PBE scheme. The 'stress-strain' method supported by CASTEP module was employed to calculate the stiffness constants.

### 3. Results and discussion

*3.1 Structural properties*

The boron-rich compounds, $B_6X$ (X = S, Se), belong to orthorhombic crystal system with space group (*Pmna*, No. 53). The studied compounds contain 28 atoms in the unit cell in which 24 are boron atoms and four are S/Se atoms. It is reported that all the boron atoms constitute $B_{12}$ clusters as icosahedral polyhedral form where S/Se atoms are at different polar and equatorial positions in the unit cell [6]. The equilibrium crystal structure (three and two dimensional) of the compound, $B_6S$ as prototype phase is depicted in Fig. 1. The predicted crystallographic lattice parameters at ambient pressure and the percentage of deviation from experimental data are collected in Table 1 along with available reported results. The deviations of lattice parameters for both boron-rich sulfide and selenide compounds are very low (~ < 0.2 %). Thus, accuracy of the present calculations is well justified.

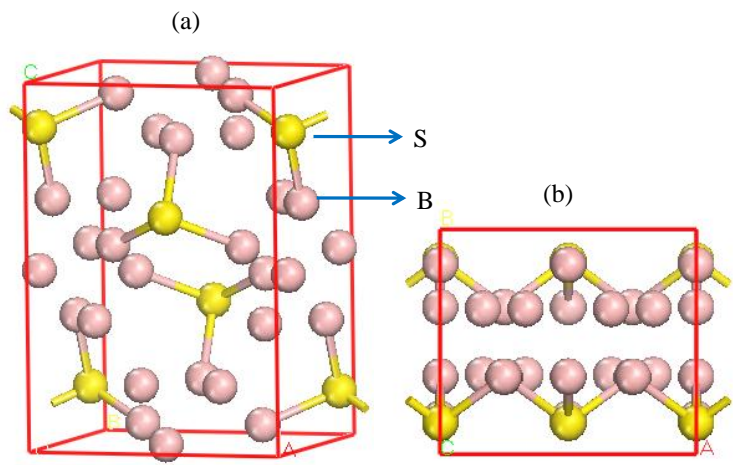

Fig. 1: Novel boron-rich chalcogenide of $B_6S$: (a) three dimensional crystal structure and (b) two dimensional structure. It should be noted that $B_6Se$ is completely isostructural.

**Table 1:** Fully relaxed lattice parameters, *a*, *b* and *c* (all in Å), unit cell volume $V$ (Å$^3$) of boron-rich chalcogenide, $B_6X$ (X = S, Se) compounds along with percentage of deviation from available experimental data.

| Phase | *a* | % of deviation | *b* | % of deviation | *c* | % of deviation | V | Code/Method | Ref. |
|---|---|---|---|---|---|---|---|---|---|
| $B_6S$ | 5.8174 | 0.0069 | 5.3077 | 0.098 | 8.1900 | 0.286 | 252.88 | CASTEP | This |
|  | 5.8170 |  | 5.3025 |  | 8.2135 |  | 253.34 | Exp. | [6] |
|  | 5.8307 |  | 5.3202 |  | 8.2072 |  | 254.59 | VASP | [6] |
|  | 5.8139 |  | 5.2918 |  | 8.2026 |  | 252.36 | CRYSTAL17 | [6] |
| $B_6Se$ | 5.9352 | 0.187 | 5.3562 | 0.032 | 8.3377 | 0.533 | 265.06 | CASTEP | This |
|  | 5.9463 |  | 5.3579 |  | 8.3824 |  | 267.06 | Exp. | [6] |
|  | 5.9684 |  | 5.3802 |  | 8.3809 |  | 269.12 | VASP | [6] |
|  | 5.9359 |  | 5.3416 |  | 8.3631 |  | 265.17 | CRYSTAL17 | [6] |

3.2 *Electronic properties, electron density difference and Mulliken population analysis*

To disclose the electronic behavior and the nature of chemical bonding, the band structure profile, electronic energy density of states (DOS), Mulliken population as well as charge density distribution of $B_6X$ (X = S, Se) have been studied for the first time. Due to lack of prior report, comparison of the present results is not possible. CASTEP code allows investigating the electronic properties using local and non-local exchange-correlations functionals. In most cases of semiconducting materials, the functionals of LDA-CAPZ and GGA-PBE underestimate the band gap [20,21]. It is reported that the Heyd-Scuseria-Ernzerhof hybrid functional (HSE06) is one of the approaches that is used to calculate a more accurate band gap of semiconducting materials [22–25]. Therefore, we have tried to explore the band gap using HSE06 along with LDA-CAPZ and GGA-PBE. Table 2 displays the estimated band gaps of both the studied compounds using LDA-CAPZ, GGA-PBE and HSE06 and it is confirmed that the band gap varies significantly for different functionals. The results presented in Table 2 need experimental verification. The electronic band structure and density of states (DOS) of $B_6X$ (X = S, Se) compounds are studied using HSE06 functional and are depicted in Figs. 2. The Fermi level ($E_F$) is set at zero energy. The bottom of the conduction band (BCB) and top of the valence band (TVB) are not at same energy as illustrated in Figs. 2. This is a clear indication of indirect band gap semiconducting behavior of these materials. The low energy part of the conduction band is caused mainly due to the strong hybridization between $p$ orbital electrons of B and S/Se elements of both compounds. Similarly, the highest of the valence band (near the Fermi level) is also comprised by the hybridization of $p$ orbitals of B and S/Se elements. It is seen that the band gap for the $B_6Se$ chalcognide is slightly higher than that of $B_6S$ as the BCB at R-point is shifted to higher energy owing to the hybridization of $p$ orbitals of B and Se elements while the TVB for both chalcogenides is found to be fixed at the G point of the Brillouin zone.

The total and partial DOS have also been studied to assess the orbital contributions to atomic bonding. In the PDOS Figs. 2 (c) and (d) of $B_6S$ ($B_6Se$) compound, the highest peak for B (B) and S (Se) is found at 5.53 (5.61) and 4.26 (4.38) eV energy, which yield the conduction band as shown in TDOS profile owing to the hybridization of these bands. The bands close to $E_F$ are comprised by $s$ and $p$ orbitals of the both compounds.

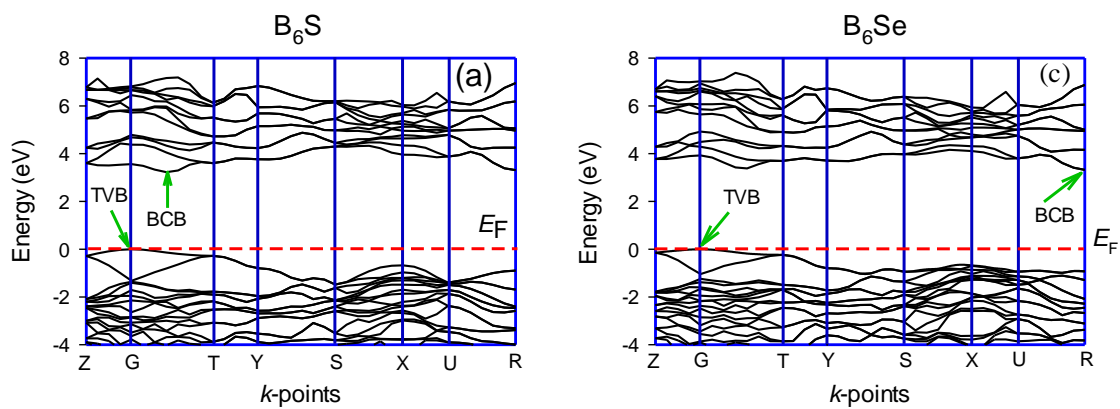

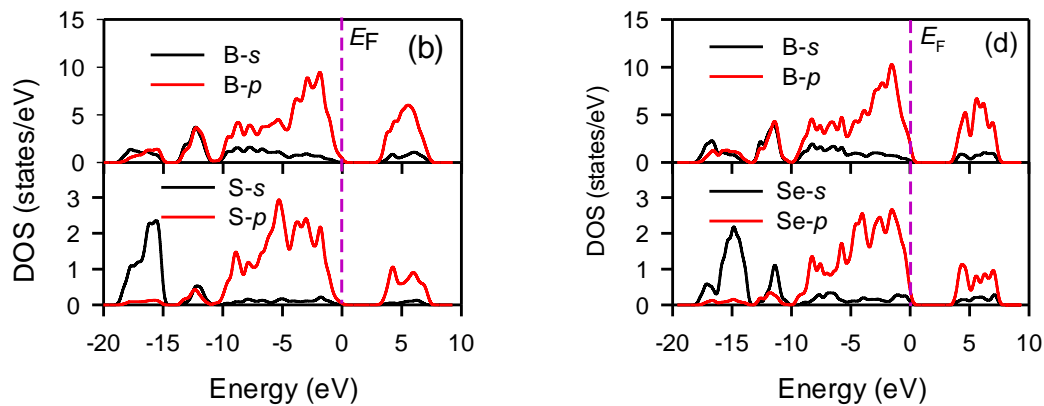

Fig. 2: Calculated (a) electronic band staructure and (b) partial density of states for $B_6S$ compound. (c-d) Same parameters for $B_6Se$ compound.

**Table 2:** Estimated band gaps of novel boron-rich sulfide and selenide compounds.

| Compound | Bandgap (eV) | Approach | Ref. |
| --- | --- | --- | --- |
| $B_6S$ | 2.789 | LDA-CAPZ | This |
|  | 2.994 | GGA-PBE | This |
|  | 3.229 | HSE06 | This |
| $B_6Se$ | 2.906 | LDA-CAPZ | This |
|  | 3.088 | GGA-PBE | This |
|  | 3.325 | HSE06 | This |

The electron density difference (EDD) with respect to the sum of the atomic densities is also studied. EDD describes the variations in the electronic charge distribution owing to the formation of all the bonds within the crystal. We have studied the EDD and Mulliken atomic population (MAP) analysis between different atomic elements of these compounds, which are crucially important in order to understand charge transfer, bonding and its nature as well [26, 27]. Fig. 3 shows the 3D visualization of EDD of $B_6X$ (X = S, Se) compounds. It is clear to see that there is strong depletion of electron density around S/Se atomic species (indicated by blue color) while accumulation of charge (red color) is found around B element in both the compounds. This suggests that charge is transferred from the S/Se to B atoms. This is indicative of ionic bonding. At the same time, strong and directional accumulation of charge is found between B – B and B – S/Se atoms which point towards formation of strong covalent bondings. The results of analysis of MAP are tabulated in Table 3. The B atoms contain negative charge while S/Se atom is indicated by positive charge. In $B_6S$ compound, donation of charge from S to B atoms as $0.02|e|$, 0.03|e|, 0.05|e| and 0.06|e| and that from Se to B atoms as 0.11|e|, 0.12|e|, 0.19|e| and 0.19|e| in B$_6$Se compound (see Table 3). These results attribute to further validation of the electron density difference analysis.

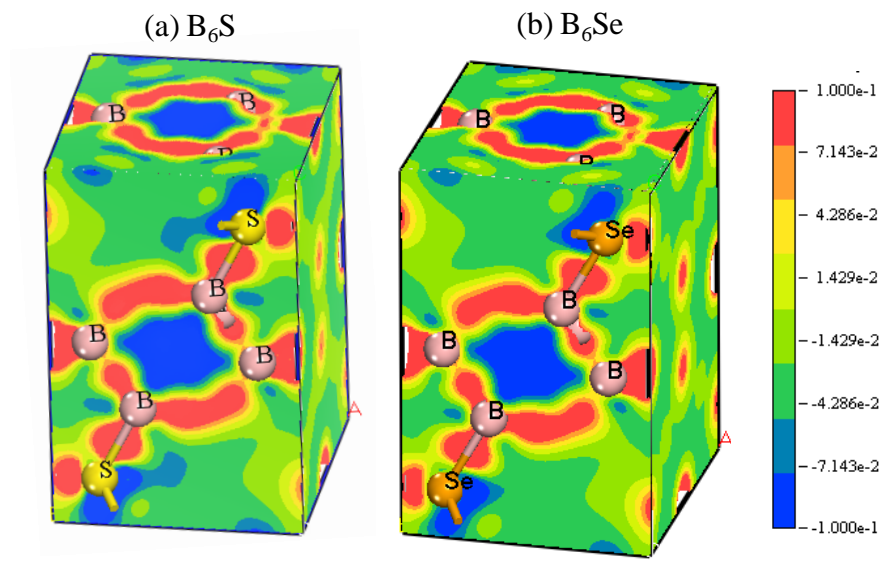

Fig. 3: Three dimensional view of electron density difference for (a) B$_6$S and (b) B$_6$Se compounds.

**Table 3:** Calculated Mulliken atomic population of boron-rich chalcogenide, B$_6$X (X = S, Se) compounds. Here, EVC indicates effective valence charge.

| Compound | Atom | s | p | Total | Charge (e) | EVC (e) |
|---|---|---|---|---|---|---|
| B$_6$S | B | 0.84 | 2.18 | 3.02 | -0.02 | --- |
|  | B | 0.80 | 2.26 | 3.06 | -0.06 | --- |
|  | B | 0.83 | 2.20 | 3.03 | -0.03 | --- |
|  | B | 0.79 | 2.26 | 3.05 | -0.05 | --- |
|  | S | 1.72 | 4.05 | 5.76 | 0.24 | 5.76 |
| B$_6$Se | B | 0.88 | 2.23 | 3.11 | -0.11 | --- |
|  | B | 0.86 | 2.33 | 3.19 | -0.19 | --- |
|  | B | 0.87 | 2.25 | 3.12 | -0.12 | --- |
|  | B | 0.85 | 2.34 | 3.19 | -0.19 | --- |
|  | Se | 1.16 | 3.93 | 5.09 | 0.91 | 5.09 |

### 3.3 *Optical spectra*

In the present section, we have examined the energy/frequency dependence of various optical parameters which are important to envisage the suitability of possible device applications of the materials under investigation. For instance, the refractive index and absorption spectra of a semiconductor are vital to design a heterostructure laser and other wave-guiding instrument [28]. Details description of theory and related formulae can be found elsewhere [29,30]. It is noted that the optical spectra for both chalcogenides are almost similar. There is only slight difference in peak positions and their intensities. The real ($\varepsilon_1(\omega)$) and imaginary ($\varepsilon_2(\omega)$) part of the dielectric function of novel $B_6S$ and $B_6Se$ compounds are displayed in Fig. 4.

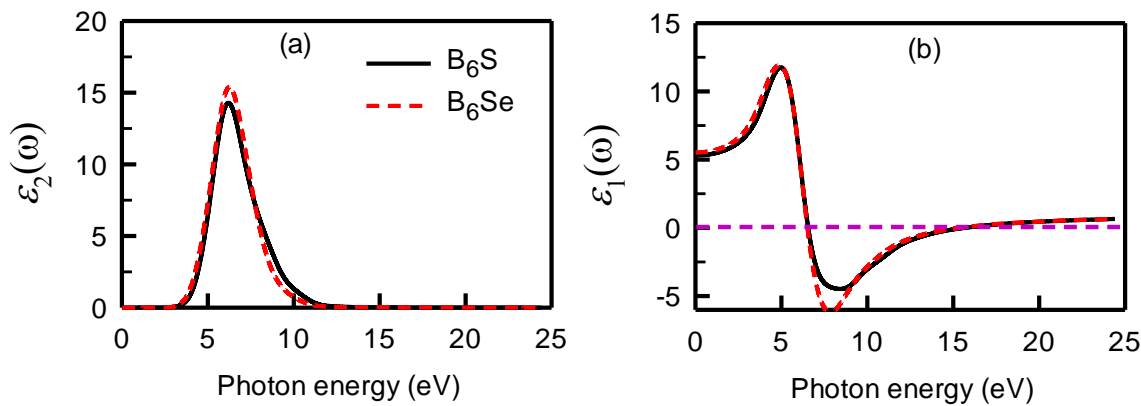

Fig. 4: The (a) imaginary part, $\varepsilon_2(\omega)$ and (b) real part, $\varepsilon_1(\omega)$ of the dielectric functions for $B_6S$ and $B_6Se$ compounds as a function of photon energy.

The photon energy dependent imaginary part of the dielectric function, $\varepsilon_2(\omega)$ could be discussed based on the band structure and PDOS analysis [27]. The $\varepsilon_2(\omega)$ usually describes the absorption phenomena of solids while the refractive index is indicated by the $\varepsilon_1(\omega)$. In Fig. 4(a), the values of $\varepsilon_2(\omega)$ for boron-rich sulfide and selenide compounds starts to increase with photon energies equal to or greater than the band gap values of 3.229 and 3.325 eV, respectively. The major peaks of $\varepsilon_2(\omega)$ are found to be at around ~6.4 eV energy for the both the compounds as shown in Fig. 4 (a). The origin of this peak is the optical transition from B-*p* orbital (high energy of VB) to S-*p* orbital (low energy of CB) due to absorption of photon with an energy ~6.4 eV. For both compounds, the static part of $\varepsilon_1(0)$ starts at around 5 as depicted in Fig. 4 (b). With the increase of photon energy, the function $\varepsilon_1(\omega)$ is gradually increased and then goes to the negative value, finally approaches zero from below at ~15 eV energy. The major peak in the real part is obtained at ~ 5 eV energy of the incident photons.

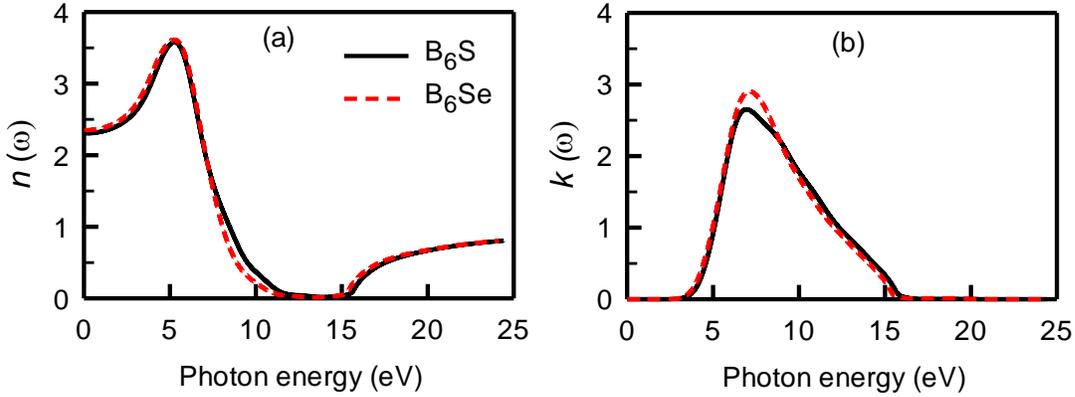

Fig. 5: Photon energy dependence of (a) refractive index, $n(\omega)$ and (b) extinction coefficient, $k(\omega)$ for $B_6S$ and $B_6Se$ compounds.

The refractive index, $n$ and extinction coefficient, $k$ are depicted in Figs. 5 (a) and (b). As we know $n$ determines the velocity of light propagation in the compound while $k$ determines its attenuation when light penetrates into a solid. The value of $n$ for both compounds is almost constant in the infrared and visible light range up to ~3.1 eV and then increases linearly to reach the peak point at 5 eV energy, after that it decreases with the increasing photon energy. The value of $k$ above the energy gap threshold is quite low. This is a consequence of relatively large band gap of $B_6S$ and $B_6Se$ compounds.

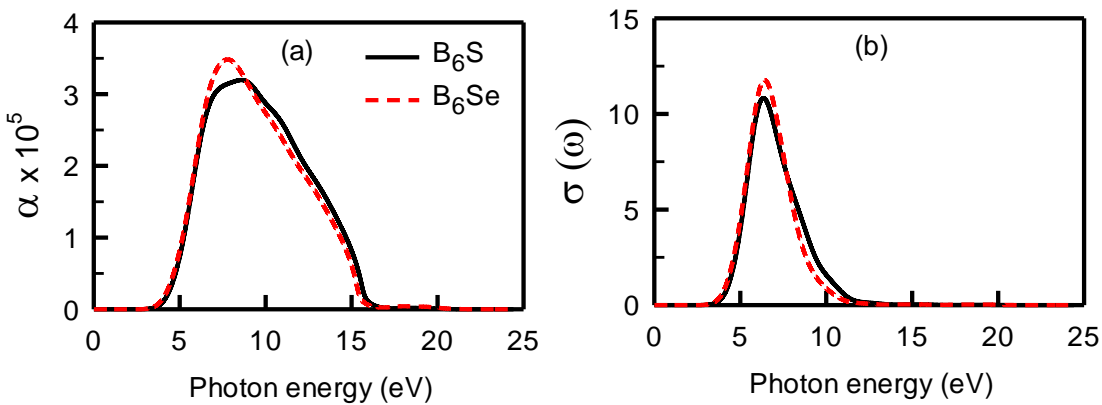

Fig. 6: Photon energy dependence of (a) absorption coefficient, $\alpha(\omega)$ and (b) photoconductivity, $\sigma(\omega)$ for $B_6S$ and $B_6Se$ compounds.

The value of the static refractive index $n(0)$ is around 2.4, which is usually low for high band gap semiconductors. Since $k$ is a measurement of absorption loss, the maximum loss appears at photon energy of ~7 eV. The peak positions for both chalcogenides are almost same, only variation is in amplitude. However, the spectral patterns of *n* and *k* consistently follow the variation of $\varepsilon_1(\omega)$ and $\varepsilon_2(\omega)$, respectively. The absorption coefficient and photoconductivity are displayed in Figs. 6. It is found that absorption edge starts at a particular finite value of photon energy, indicating that the compounds have band gaps that are also confirmed from the band structure analysis shown in Fig. 2. The photon energy dependent spectra of $\alpha(\omega)$ is shown in Fig. 6(a). It illustrates large absorption capacity mainly in the ultraviolet region, which suggests that the studied materials are not transparent, can crucially be used in UV surface-disinfection device, medical autoclave, medical sterilizer equipment and even to design some optoelectronic devices [31]. The trend of photoconductivity spectra follows the same route of $\alpha(\omega)$ as expected. We also found that the photoconductivity starts when the photon energy exceeds the band gap energy as shown in Fig. 6(b). The reflectivity spectra were also evaluated from the dielectric constant which is given in Fig. 7. The reflectivity spectra start at 0.16 (16%), show almost constant value in the visible light region and then increases with the highest value at around 13.5 eV where $\varepsilon_1(\omega)$ goes to zero (Fig. 4b). The reflectivity is found to be less than 20% in the visible and near-UV region (up to 4 eV). The loss function, $L(\omega)$, is associated with the energy loss of fast electrons in a material, which is shown in Fig. 7 (b). Prominent energy loss is seen in the mid-UV region (at 15.4 eV for both compounds) and the peak appeared owing to the plasma resonance. The Plasmon energy (peak energy of $L(\omega)$) can be correlated with $R(\omega)$ where the reflectivity falls sharply from high value and $\varepsilon_1(\omega)$ approaches zero as shown in Figs. 4(b) and 7 (a). It should be noted that the validation of the studied optical properties is not possible as there is no prior relevant report on these compounds experimentally and/or theoretically.

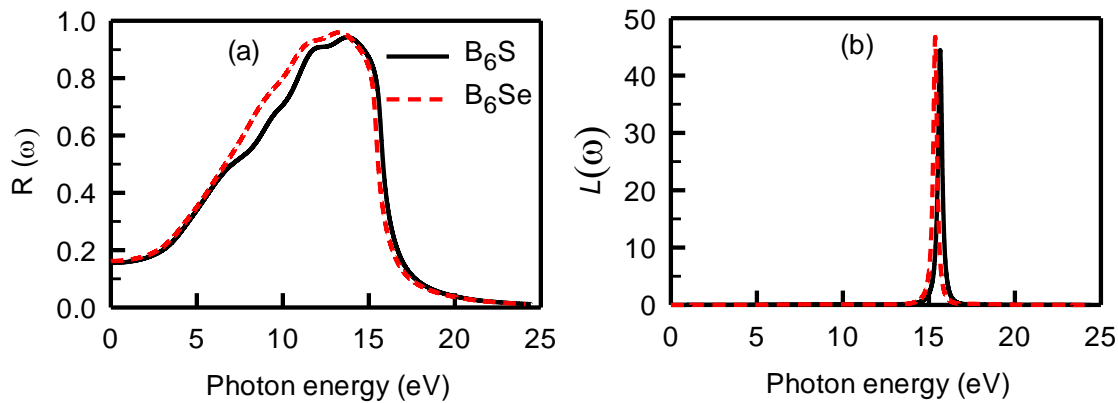

Fig. 7: Photon energy dependence of (a) reflection coefficient, $R(\omega)$ and (b) loss function, $L(\omega)$ of $B_6S$ and $B_6Se$ compounds.

We believe that the prediction of optical spectra herein would motivate researchers to investigate the optoelectronic properties of these compounds further for possible practical device applications.

3.4 *Stiffness constants and mechanical properties*

Study of mechanical properties of crystalline solids helps to understand its response under the influence of applied loads of different nature. It also assists to predict their possible applications in many sectors by exploring physical properties such as stiffness, fracture toughness, mechanical stability, elastic moduli, ductility, brittleness, hardness and elastic anisotropy. We have organized the study of mechanical properties as follows: firstly, we have calculated the independent elastic constants within the strain–stress method [21,26,32,33]. The obtained elastic constants are then used to calculate the polycrystalline elastic moduli using Hill's approximation that represents the average values of upper and lower bounds of the elastic moduli [34]. Both elastic constants and elastic moduli are then used to evaluate the hardness, fracture toughness, ductility/brittleness and anisotropic behavior for the considered chalcogenide systems. Table 4 shows the obtained elastic constants of $B_6S$ and $B_6Se$. The mechanical stability criteria of orthorhombic structures are presented as follows [33,35]:

$C_{11} > 0; C_{11}C_{22} > C_{12}^2; C_{11}C_{22}C_{33} + 2C_{12}C_{13}C_{23} - C_{11}C_{23}^2 - C_{22}C_{13}^2 - C_{33}C_{12}^2 > 0, C_{44} > 0; C_{55} > 0; C_{66} > 0.$

**Table 4:** Calculated stiffness constants (GPa) of boron-rich chalcogenide, $B_6X$ (X = S, Se) compounds with orthorhombic structure.

| Compound | $C_{11}$ | $C_{22}$ | $C_{33}$ | $C_{44}$ | $C_{55}$ | $C_{66}$ | $C_{12}$ | $C_{13}$ | $C_{23}$ | Ref. |
|---|---|---|---|---|---|---|---|---|---|---|
| $B_6S$ | 417 | 277 | 463 | 115 | 203 | 78 | 2 | 75 | 32 | This |
| $B_6Se$ | 377 | 275 | 450 | 137 | 182 | 90 | 5 | 60 | 48 | This |
| $B_6Si$ | 205 | 338 | 397 | 100 | 123 | 72 | 79 | 97 | 57 | [11] |

As seen from the Table 4, the elastic constants of both $B_6S$ and $B_6Se$ satisfy the aforementioned conditions, indicating the mechanical stability of these compounds. In general, the values of $C_{11}$, $C_{22}$ and $C_{33}$ are larger than that of $C_{44}$ for mechanically stable phases. For $B_6S$ and $B_6Se$, the $C_{ij}$ (i = j = 1/2/3) are larger than $C_{44}$ which hints for comparatively higher incompressibility along *a*-, *b*- and *c*-axis. Moreover, $C_{33} > C_{11} > C_{22}$ for both the compounds, the unequal values are due to the differences in the atomic arrangement along *a*-, *b*- and *c*-axis. The atoms are arranged in different pattern along different axis to minimize the total energy of the crystal structure in the ground state. The highest values of $C_{33}$ compared to $C_{11}$ and $C_{22}$ confirms that the atomic bonding along *c*-direction is stronger than those along the *a*- and *b*-directions in the crystal. This condition also informs that the lattice constant *c* is less sensitive to pressure and temperature compared to *a* and *b*. The values of $C_{11}$, $C_{22}$ and $C_{33}$ are larger than those of $C_{12}$, $C_{13}$ and $C_{23}$, indicating stronger resistance to deformation along axial direction compared to shear deformations. The unequal values are also responsible for the elastic anisotropy possessed by the systems under consideration. So far we know, there is no experimental or theoretical data for the elastic constants of B$_6$S and B$_6$Se that can be used to compare our results. Thus, the results obtained in this study could be used as reference for future studies. We have listed the elastic constants of B$_6$Si in Table 4 for comparison [11].

**Table 5:** Calculated elastic moduli (*B, G* & *Y*), Poisson's ratio (υ), Pugh's ratio (G/B) hardness values (H$_{micro}$ and H$_{macro}$) and fracture toughness ($K_{IC}$) of boron-rich chalcogenide, B$_6$X (X = S, Se) compounds.

| Compound | B (GPa) | G (GPa) | Y (GPa) | υ | G/B | B/C$_{44}$ | H$_{micro}$ (GPa) | H$_{macro}$ (GPa) | $K_{IC}$ (MPam$^{0.5}$) | Ref. |
|---|---|---|---|---|---|---|---|---|---|---|
| B$_6$S | 147 | 140 | 319 | 0.14 | 0.95 | 1.28 | 33.60 | 31.02 | 2.070 | This |
|  | 146 | 138 | 315 | 0.14 | 0.94 |  |  | 31 | 2.1 | [6] |
| B$_6$Se | 143 | 142 | 320 | 0.13 | 0.99 | 1.04 | 35.57 | 33.02 | 2.072 | This |
|  | 137 | 135 | 304 | 0.13 | 0.98 |  |  | 32 | 2.0 | [6] |

The bulk modulus *B* and shear modulus *G* can be estimated using the elastic constants. The *B* and *G* are calclated based on the Hill's approximation [34] which calculates the actual elastic moduli by averaging the upper bound (Voigt bounds [36]) and lower bounds (Reuss bounds [37]) of the elastic moduli. For the orthorhombic crystals, the Voigt bounds of bulk ($B_V$) and shear ($G_V$) moduli are calculated using the following equations [36]:

$$B_V = \frac{1}{9}[C_{11} + C_{12} + C_{33} + 2(C_{12} + C_{13} + C_{23})]$$

$$G_V = \frac{1}{15}[C_{11} - C_{12} - C_{13} + C_{22} - C_{23} + C_{33} + 3(C_{44} + C_{55} + C_{66})]$$

For the orthorhombic crystals, the Reuss bounds of bulk ($B_R$) and shear ($G_R$) moduli are calculated using the following equations [37]:

$$B_R = \chi[C_{11}(C_{22} + C_{33} - 2C_{23}) + C_{22}(C_{33} - 2C_{13}) - 2C_{33}C_{12} + C_{12}(2C_{23} - C_{12}) + C_{13}(2C_{12} - C_{13}) + C_{23}(2C_{13} - C_{23})]^{-1}$$

$$G_R = 15\{4\,[C_{11}(C_{22} + C_{23} + C_{33}) + C_{22}(C_{33} + C_{13} + C_{33}C_{12} - C_{12}(C_{23} + C_{12}) - C_{13}(C_{12} + C_{13}) - C_{23}(C_{13} + C_{23})]/\chi + 3(C_{44}^{-1} + C_{55}^{-1} + C_{66}^{-1})\}^{-1}$$

$$\chi = C_{13}(C_{12}C_{23} - C_{13}C_{22}) + C_{23}(C_{12}C_{13} - C_{23}C_{11}) + C_{33}(C_{11}C_{22} - C_{12}^2)$$

Hill's average [38,39] of $B$ and $G$ are calculated by: $B = (B_V + B_R)/2$ and $G = (G_V + G_R)/2$. The obtained values of $B$ and $G$ are then used to calculate the Young's modulus and Poisson's ration as follows: $Y = 9BG/(3B + G)$ and $v = (3B - Y)/(6B)$ [40,41]. The estimated values are presented in Table 5.

The value $B$ signifies the resistance to a change in the volume of a crystal while the resistance to a change in the shape is signified by $G$. The values of $Y$ are indicators of stiffness of solids. The elastic moduli such as $B$, $G$ and $Y$ are not only helpful to know the mechanical properties solids but also useful to evaluate the hardness of materials. Among the elastic constants $C_{44}$ is the best one to predict the hardness of solids [42]. As evident from the listed values in Table 4, the $B_6Se$ is harder than $B_6S$. On the other hand, shear modulus ($G$) is the best hardness predictor among the elastic moduli which is also found in good agreement with previous statement. These statements can be proved by the calculation of hardness parameters $H_{macro}$ and $H_{micro}$ where the formulae can be expressed as: $H_{macro} = 2[(\frac{G}{B})^2 G]^{0.585} - 3$ [43] and $H_{micro} = \frac{(1-2v)Y}{6(1+v)}$ [44]. The listed values are presented in Table 5 and found to be consistent with the predictions based on the values of $C_{44}$ and $G$. The obtained elastic moduli and hardness, $H_{macro}$ (using Chen's formula) are in very good agreement with the previous results [6]. However, $H_{micro}$ is observed to be higher than that of $H_{macro}$ obtained using Chen's formula. The difference in the hardness values comes due to the parameters involved in the equations. Cherednichenko et al., [6] have predicted the studied compounds to be the members of hard materials based on the calculated hardness using Chen formula [45]. The materials are said to be super hard with hardness value greater than 40 GPa. The hardest material known so far is the diamond with hardness (using Chen's formula) 93.6 GPa. Mazhnik et al., [46] have calculated the hardness of different materials using Chen's formula and found to have a very good agreement between the theoretical and experimental values. The hardness values of $B_6S$ and $B_6Se$ are lower than those of $BC_2N$ (76.5 GPa), $BC_5$ (63.6 GPa), c-BN (62.4 GPa), $\gamma$-$B_{28}$ (49.0 GPa) [46]. The elastic moduli of the above mentioned compounds have been calculated by density functional theory using the generalized gradient approximation (GGA) of the Perdew–Burke–Ernzerhof (PBE) [47] as exchange-correlations terms. We have also used the same formalism but calculations code is different (CASTEP for us and VASP for [46]). It is well known, Diamond, $BC_2N$, $BC_5$, c-BN, $\gamma$-$B_{28}$ are widely used super hard materials [46]. The much lower hardness of $B_6S$ (31 GPa) and $B_6Se$ (33 GPa) than that of the super hard materials confirms that the studied compounds are not super-hard materials. Now, let us have a look on some other hard materials. Mazhnik et al., [46] have also calculated the hardness of $B_4C$ (32.6), $B_6O$ (35.5 GPa), $\beta$-SiC (34.8 GPa), $SiO_2$ (30.0 GPa), WC (33.5 GPa), $OsB_2$ (17.8 GPa), VC (26.5 GPa), $Re_2B$ (38.6 GPa) etc. These materials are widely classified as hard materials. It should be noted again that these calculated values are consistent with the experimental values and appeals to be accepted them as references. Though there is no report on experimental hardness of the $B_6S$ and $B_6Se$ but our obtained values are in very good agreement with reported values by Cherednichenko et al., [6] where they have used the VASP code with the generalized gradient approximation (GGA) of the Perdew–Burke–Ernzerhof (PBE)as exchange-correlations terms [13,18]. Thus it is reasonable to expect that the experimental hardness value of the $B_6S$ and $B_6Se$ compounds would be close to the calculated values in this work. Based on the above discussion, we claim that the titled compounds $B_6S$ and $B_6Se$ are to be categorized as hard materials as found by Cherednichenko et al. [6].

Furthermore, the fact that estimated hardness ( micro and macro) values of $B_6Se$ are slightly higher than those of $B_6S$ can be explained using the density of states (DOS) profiles of both the compounds. The total DOS of both compounds is shown in Fig. 8, where a slight shift of the peaks in energy is noticed (indicated by red lines) towards low energy side from the Fermi level. The peaks observed in the DOSs resulted from the hybridization between the electronic states of B and S/Se atoms, leading in the formation of strong covalent bonds between them. The bonding strength depends on the energy level of the peaks; lower the energy level, stronger is the covalent bonds. This results in higher values of hardness parameters for $B_6Se$ than those for $B_6S$. Similar results were also been observed for other ceramic materials [39–42].

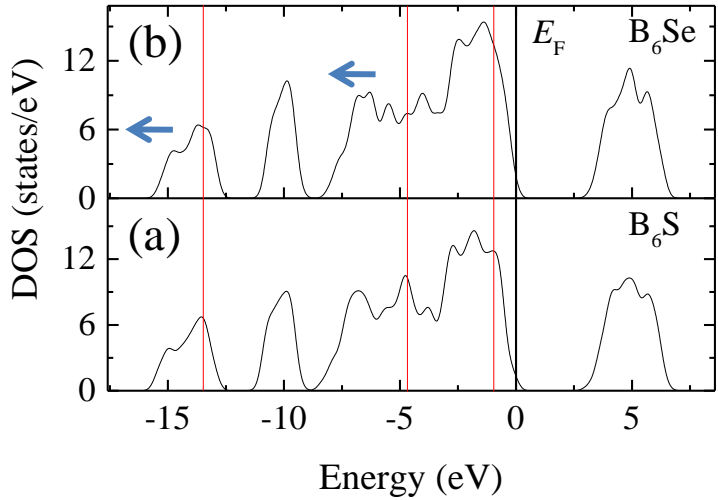

Fig. 8: Total density of states of $B_6S$ and $B_6Se$ compounds. The black and red vertical lines indicate the Fermi level ($E_F$) set to zero and peak positions, respectively (please consult text for details).

Along with the hardness, the fracture toughness ($K_{IC}$) is an extremely important parameter that characterizes the mechanical properties of solids [51]. It provides the resistance of a solid that is able to prevent the propagation of a crack produced inside. A material should have high fracture toughness along with hardness for the industrial applications, consequently prediction of the fracture toughness has drawn significant engineering interest [52–54]. The possibility of $B_6S$ and $B_6Se$ to be used as hard materials, compels us to calculate the $K_{IC}$ via the formula $K_{IC} = V_0^{1/6} \cdot G \cdot (B/G)^{1/2}$ [46]. The obtained values of $K_{IC}$ for the $B_6S$ and $B_6Se$ phases are found to be 2.070 MPam$^{0.5}$ and 2.072 MPam$^{0.5}$, respectively. These are lower than those of super hard diamond (6.3 MPam$^{0.5}$), WC (5.4 MPam$^{0.5}$) and c-BN (5.4 MPam$^{0.5}$) as reported by Mazhnik et al. [46].

A challenge for the application of hard materials is its machinability. The machinability index (MI) can be evaluated from the ratio of $B/C_{44}$. Though, the obtained values of MI for considered solids are low due to their high hardness values but are comparable to the MI of well known machinable MAX phase materials [31,48–50,55]. Thus, the $B_6S$ and $B_6Se$ are expected to possess moderate level of machinability.

The values of *B* and *G* are further used to assess the brittleness of $B_6S$ and $B_6Se$. As proposed by Pugh, the ratio of *G/B* is useful to predict brittle or ductile behavior of a solid [56]. A solid is said to behave as ductile when the value of *G/B* is smaller than 0.57, otherwise it should be brittle. As evident from the values presented in Table 5, the studied compounds are brittle in nature. The *B* and *Y* values are used to calculate the Poisson's ratio (υ) that is also used to predict the ductile/brittle behavior of solids. A value of 0.26 is used as a separating value; a higher value ( > 0.26) is set for ductile material and lower value (< 0.26) is set for brittle materials [48,49]. As seen in Table 5, the compounds studied herein are brittle materials. In addition, information regarding chemical bonding within the solids can also be known from the values of υ. The value of υ is typically small (υ = 0.10) for covalent solids whereas υ is 0.33 for ionic solids [57]. The obtained values of υ (0.14 and 0.13) indicate that the chemical bonding within $B_6S$ and $B_6Se$ are largely of covalent type. Furthermore, the dominant inter-atomic forces within the solids are central inter-atomic forces when υ is in between 0.25 to 0.50, otherwise the inter-atomic forces are non-central within the solids. The obtained values do not fall into the aforementioned range, revealing that the non-central inter-atomic forces are dominant within $B_6S$ and $B_6Se$. Pettifor [58] proposed another parameter that can be used to understand the nature of chemical bonding as well as ductile/brittle nature of solids based on the value of Cauchy pressure (*CP*). The *CP* can be obtained from the elastic constant using the relations: $CP = C_{23}–C_{44}$, $C_{13}–C_{55}$ and $C_{12}–C_{66}$. The *CP* will be lower than zero for the brittle solid whiles a value higher than zero implies the ductile solid [59]. The negative values of *CP*s [*CP* of $B_6S$ ($B_6Se$) are -83 (-89), -128 (-122) and -76 GPa (-85 GPa)] reveal that both compounds are brittle in nature. Moreover, the large negative value of CP is considered as an indicator for the directionality in the covalent bonds [33]. Thus, the large negative values of *CP*s for both the studied compounds indicate considerable directional covalent bonding within their structure.

## 3.5 *Elastic anisotropy*

Elastic anisotropy is one of the key properties of materials that reveal the difference in the atomic arrangement in the different directions. Moreover, a proper description of the anisotropic nature of solids helps to explore the mechanical stability by assessing the formation and propagation of micro-cracks inside materials. It also helps to enhance the parameters that characterize the mechanical strengths of solids [60,61]. Thus study of elastic anisotropy of solids has notable importance in materials engineering and in crystal physics.

There are a number of different anisotropy parameters characterizing the mechanical anisotropy of solids with respect to different types of stresses. For the orthorhombic solids like $B_6S$ and $B_6Se$, the shear anisotropic factors : $A_{\{100\}}$, $A_{\{010\}}$ and $A_{\{001\}}$ for the {100}, {010} and {001} planes, respectively can be calculated by the formulae:

$$A_{\{100\}} = 4C_{44}/(C_{11} + C_{33} - 2C_{13}); A_{\{010\}} = 4C_{55}/(C_{22} + C_{33} - 2C_{23});$$

$$A_{\{001\}} = 4C_{66}/(C_{11} + C_{22} - 2C_{12})$$ [62]. The value of $A_{\{100\}}$, $A_{\{010\}}$ and $A_{\{001\}}$ equal to one implies the isotropic nature of solids with respect to shear along the principle crystal planes while other values give the extent of shear anisotropy. The obtained values of the $B_6S$ and $B_6Se$ indicating the anisotropic nature of these to boron-rich compounds.

The percentage anisotropy in compressibility ($A_B$) and shear ($A_G$) is defined as $A_B = \frac{B_V - B_R}{B_V + B_R}$ and $A_G = \frac{G_V - G_R}{G_V + G_R}$ [63]. A value of zero and 100% for $A_B$ and $A_G$ implies the isotropy and maximum anisotropy of crystals, respectively. Slightly anisotropic nature of $B_6S$ and $B_6Se$ is revealed by the low values of $A_B$ and $A_G$.

The universal anisotropy index $A^U$ [64] can be obtained based on the upper limit (Voigt, *V*) and lower limit (Reuss, *R*) of bulk and shear modulus using the relation: $A^U = 5\frac{G_V}{G_R} + \frac{B_V}{B_R} - 6 \geq 0$. A non-zero value of $A^U$ is indicative of mechanical anisotropy. Different anisotropic indices: shear anisotropic factors ($A_{100}$, $A_{010}$ and $A_{001}$), percentage anisotropy ($A_B$ and $A_G$) and universal anisotropy index $A^U$ of boron-rich chalcogenide, $B_6X$ (X = S, Se) compounds are summarized in Table 6.

**Table 6:** Different anisotropic indices: shear anisotropic factors ($A_{100}$, $A_{010}$ and $A_{001}$), percentage anisotropy ($A_B$ and $A_G$) and universal anisotropy index $A^U$ of boron-rich chalcogenide, $B_6X$ (X = S, Se) compounds.

| Compound | $A_{100}$ | $A_{010}$ | $A_{001}$ | $A_B$ | $A_G$ | $A^U$ |
|---|---|---|---|---|---|---|
| $B_6S$ | 0.63 | 1.20 | 0.45 | 3.99 | 6.84 | 0.82 |
| $B_6Se$ | 0.77 | 1.16 | 0.56 | 3.27 | 3.80 | 0.46 |

To demonstrate the elastic anisotropy in further detail, the variation of Young's modulus, compressibility, shear modulus and Poisson's ratio along different crystallographic directions should be considered. With an intention to explore this anisotropic nature we have estimated the 3D and plane-projected 2D variations of *Y, K, G* and υ of $B_6S$ and $B_6Se$ using the ELATE code [65]. The mechanical anisotropy of both compounds is similar except the slight differences in the magnitude and we have presented the plots only for $B_6S$. Fig. 9 (a) shows the 3D representation of *Y* from which the anisotropic nature is clearly observed. To distinguish the behavior in different planes, its 2D plots in three major planes are also shown in Fig. 9 (b). As evident from Fig. 9(a), the *Y* is more anisotropic in *xy* planes. The least anisotropy is observed in the *xz* plane in which the maximum values are at an angle of 45° between the vertical and horizontal axes. The extent of anisotropy for *yz* plane is midway between those in the *xy* and *xz* planes in which the maximum is on the vertical axis.

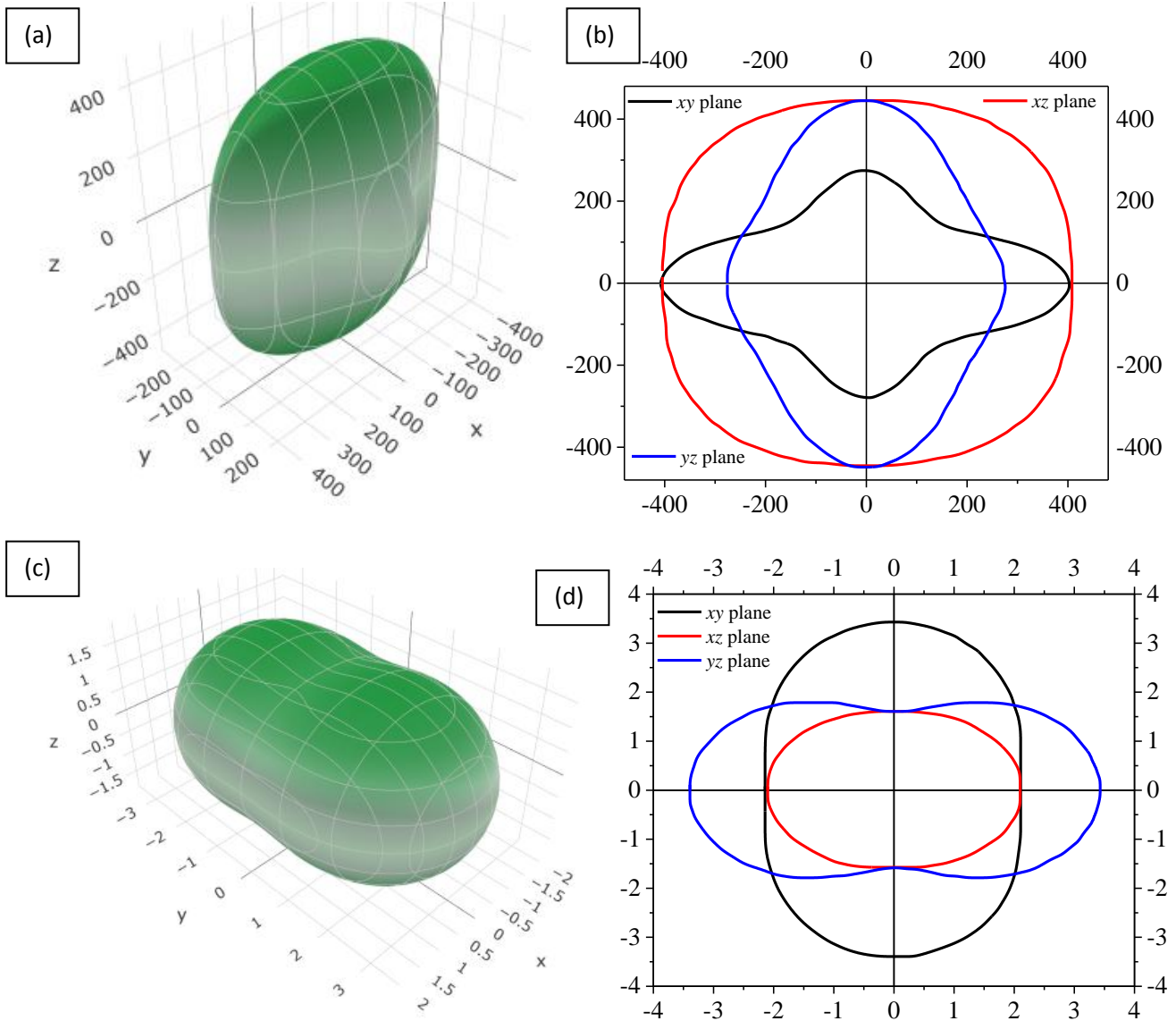

Fig. 9: (a) 3D view and (b) 2D view of Young modulus (Y); (c) 3D view and (d) 2D view of compressibility for $B_6S$.

Fig. 9(c) shows the 3D plot demonstrating the anisotropy in compressibility. The 2D anisotropic nature can be visualized in Fig. 9 (d), where the anisotropies in the three major planes are indicated. In *xy* plane, the maximum and minimum of compressibility are on the vertical axis and horizontal axis, respectively. In *xz* and *yz* plane, the maximum is on the horizontal axis and the minimum is on the vertical axis. As seen from the figure, the largest ratio of maximum to the minimum value is observed for the *yz* plane.

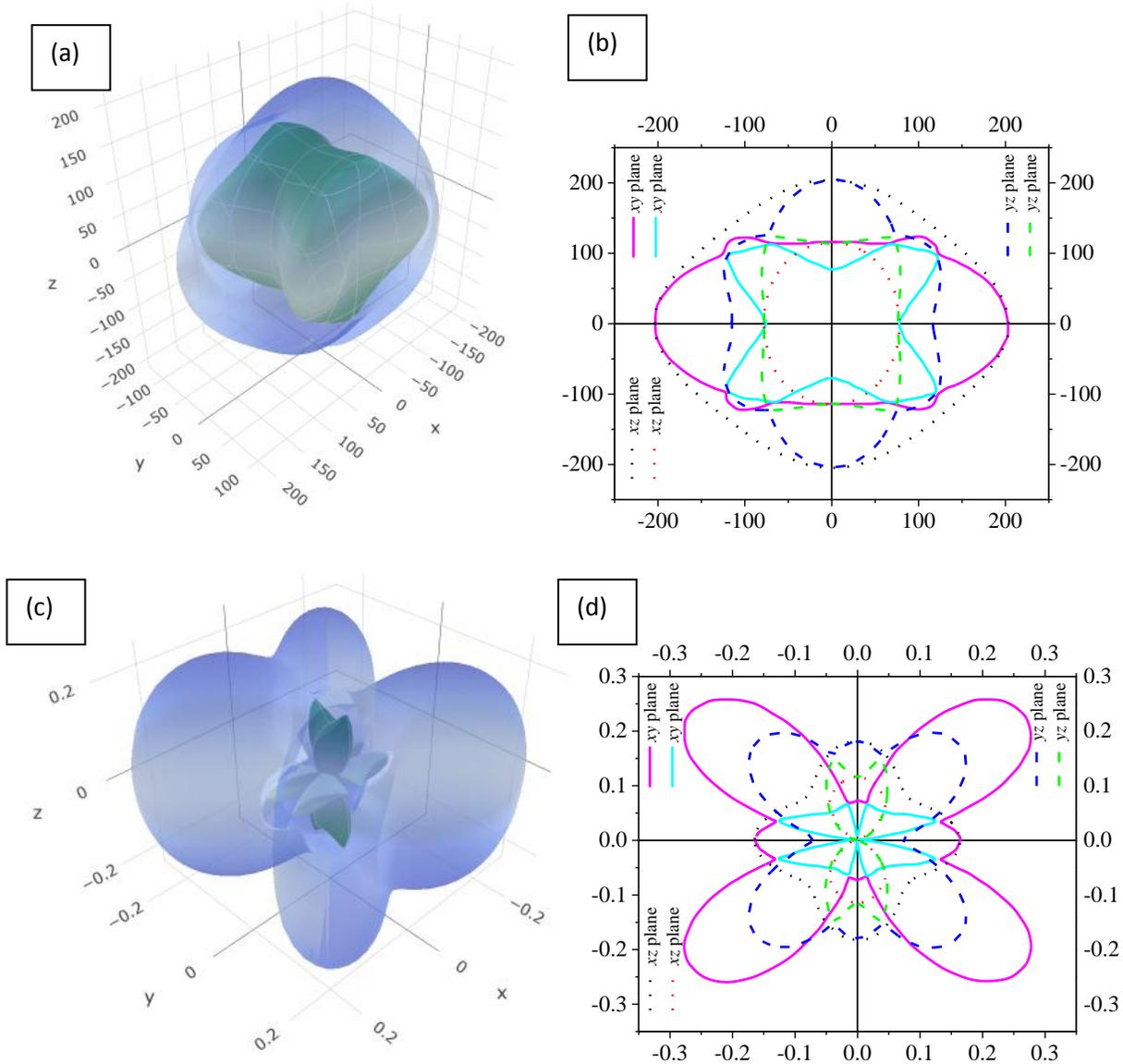

Fig. 10: (a) 3D view and (b) 2D view of shear modulus (G); (c) 3D view and (d) 2D view of Poisson's ratio (υ) for chalcogenide $B_6S$ to observe the anisotropic nature is clearly.

Unlike the Young's modulus and compressibility, the shear modulus exhibits two surfaces for each plane: outer surface exhibits maximum values at each of angle while the inner surface indicates the minimum values of the shear modulus. For the *xy* plane: the maximum value of the pink (outer surface) line is on the horizontal axis while it has minimum value on the vertical axis.

The cyan line (inner surface) has the maximum value at an angle of around 30° from the horizontal axis in all four quadrants and minimum values on the axes. For the *xz* plane, the black dotted line (outer surface) has the maximum values on the axes and minimum values at an angle of 45° from the axes. For the *yz* plane, the maximum values of the outer surface are on the vertical axis while the minimum values are at an angle 30° from the vertical axis on both sides. Like shear modulus, the Poisson's ratio also exhibits two surfaces for each plane in which complex anisotropic nature is observed. The maximum and minimum values of *Y, K, G, υ* and their maximum to minimum ratios are presented in Table 7. These ratios are useful indicators of elastic anisotropy.

**Table 7:** The minimum and the maximum values of the Young's modulus, compressibility, shear modulus, and Poisson's ratio of $B_6S$ and $B_6Se$.

| Phase | $Y_{min.}$ (GPa) | $Y_{max.}$ (GPa) | $A_Y$ | $K_{min}$ $(TPa^{-1})$ | $K_{max}$ $(TPa^{-1})$ | $A_K$ | $G_{min.}$ (GPa) | $G_{max.}$ (GPa) | $A_G$ | $υ_{min.}$ | $υ_{max.}$ |
|---|---|---|---|---|---|---|---|---|---|---|---|
| $B_6S$ | 207.88 | 458.32 | 2.20 | 1.58 | 3.41 | 2.15 | 77.63 | 203.99 | 2.63 | -0.0117 | 0.355 |
| $B_6Se$ | 225.63 | 432.09 | 1.91 | 1.55 | 3.32 | 2.13 | 90.322 | 182.45 | 2.02 | -0.0060 | 0.266 |

*3.6 Phonon dispersion curves and phonon density of states*

Dynamical stability check of solids is useful from the point of view of since materials are often subject to time-varying mechanical stress. The phonon dispersion curve (PDC) at absolute zero has been calculated using the density functional perturbation theory (DFPT) linear-response method to check the dynamical stability of the materials under consideration and vibrational contribution to the thermodynamic properties such as linear thermal expansion coefficient and heat capacities [66–68]. The PDC along with the total phonon density of states (PHDOS) of $B_6X$ (X = S, Se) compounds along the high symmetry directions of the crystal Brillouin zone (BZ) are depicted in Figs. 11 (a) and (b). The stability has been checked using the phonon frequency over the whole BZ. In general, existence of any negative frequencies at any *k*-points certifies the instability of the compounds otherwise they are considered dynamically stable. No negative frequency phonon mode is observed in the displayed PDCs in Figs. 11 (a) and (b), consequently, the compounds $B_6X$ (X = S, Se) are dynamically stable.

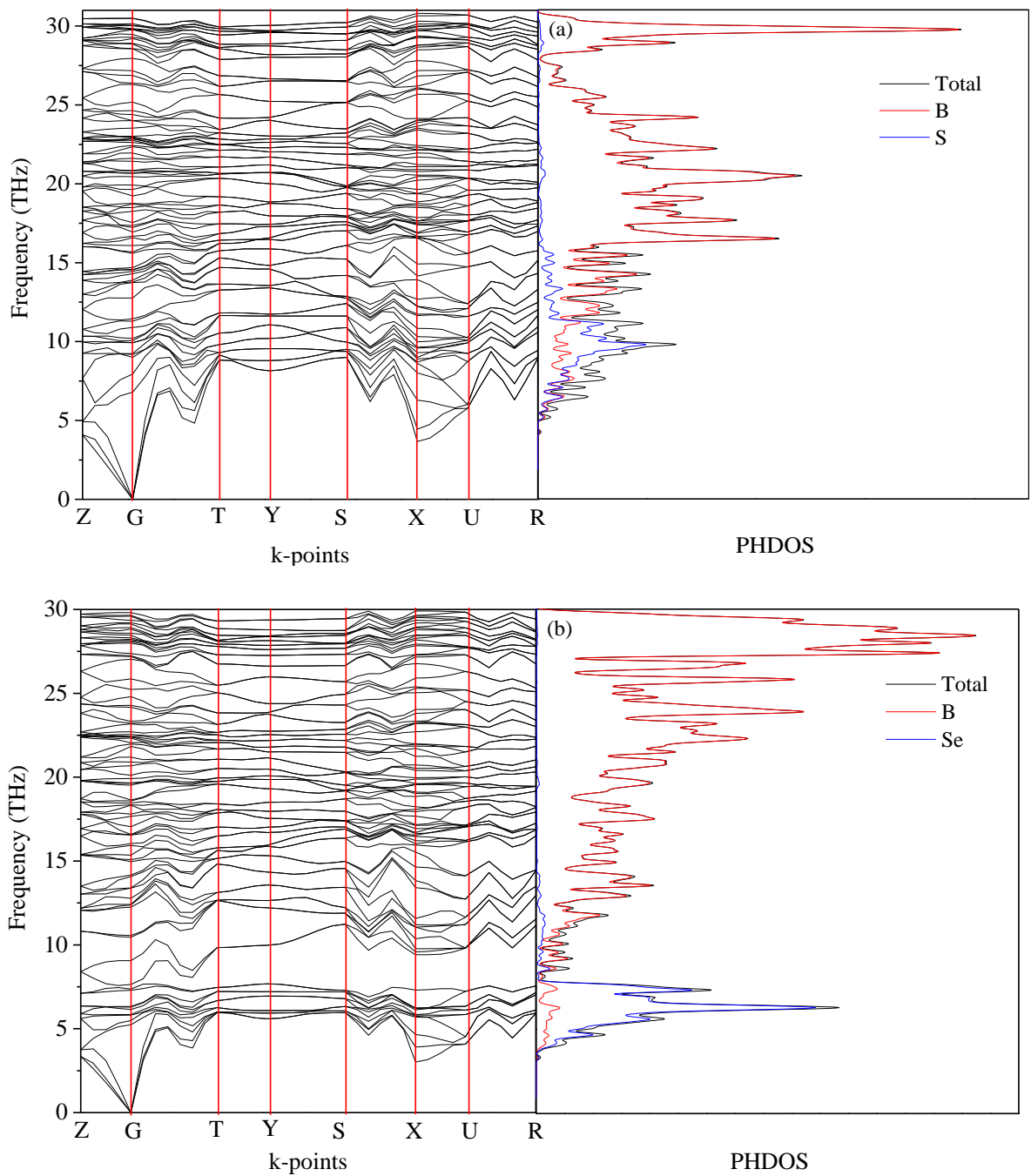

Fig. 11: The phonon dispersion curves (PDCs) and phonon density of states (PHDOS) of (a) $B_6S$ and (b) $B_6Se$ compounds.

The PHDOS of the compounds $B_6X$ (X = S, Se) are displayed side by side for better understanding in Figs. 11(a) and (b), respectively, along with PDC to identify the bands and their corresponding density of states. The prominent peaks are observed due to the flatness of the bands in both the compounds while non-flat bands result in weak peaks in the PHDOS. The phonon modes in the low frequency region (5-12 THz) of $B_6S$ compound are mainly comprised of vibrations of S atoms while those in the high frequency region (> 12 THz) are credited to the B species as shown in Fig. 11 (a). In case of $B_6Se$ compound, lower frequency region (3-8 THz) in the PDC curve is mainly composed of motion of Se atoms. The phonon spectra of B atoms for $B_6Se$ compound are mainly responsible for modes in the high frequency region (> 8 THz) as shown in Fig. 11 (b).

*3.7 Debye temperature*

The frequency of a crystal's highest normal mode of vibration is represented by the Debye frequency or equivalently by Debye temperature, $\Theta_D$. It correlates the elastic properties to the thermodynamic properties such as phonons, thermal expansion, thermal conductivity, specific heat, and lattice enthalpy. Usually, the materials with larger Debye temperature $\Theta_D$ possess high hardness and high lattice thermal conductivity. The Debye temperature can be calculated from the average sound velocity ($v_m$) which is associated to shear modulus (*G*) and bulk modulus (*B*). $\Theta_D$ can be calculated with the mean sound velocity $v_m$ which is expressed by the following equation:

$$v_m = \left[1/3\left(1/v_l^3 + 2/v_t^3\right)\right]^{-1/3}$$

where, $v_l$ and $v_t$ are longitudinal and transverse sound velocity, respectively. The $v_l$ and $v_t$ are represented by the following equations related to the elastic moduli (shear and bulk modulus) and density of the solid under consideration:

$$v_l = [(3B + 4G)/3\rho]^{1/2} \text{ and } v_t = [G/\rho]^{1/2}.$$

Finally, $\Theta_D$ can be calculated using the expression [69]:

$$\Theta_D = h/k_B \left[\left(3n/4\pi\right)N_A\rho/M\right]^{1/3} v_m,$$

where, $M$, $n$, $\rho$, $h$, $k_B$, and $N_A$ are the molar mass, the number of atoms in the molecule, the mass density, the Planck's constant, Boltzmann constant, and Avogadro's number, respectively. The vibration frequency of particles is changed with temperature that is reflected by the Debye temperature, $\Theta_D$. The calculated Debye temperature, $\Theta_D$ along with the density ($\rho$), different sound velocities ($v_l$, $v_t$ and $v_m$) and $\Theta_D$ of the compositions are summarized in Table 8.

**Table 8:** Calculated crystal density, longitudinal, transverse and average sound velocities ($v_l$, $v_t$, and $v_m$), Debye temperature, $\Theta_D$, minimum thermal conductivity, $K_{min}$, lattice thermal conductivity, $k_{ph}$ at 300 K and Grüneisen parameter, $\gamma$, for novel boron-rich sulfide and selenide compounds.

| Compounds | $\rho$ (kg/m³) | $v_l$ (km/s) | $v_t$ (km/s) | $v_m$ (km/s) | $\Theta_D$ (K) | $K_{min}$ (W/mK) | $k_{ph}^*$ (W/mK) | $\gamma$ | $T_m$ (K) |
|---|---|---|---|---|---|---|---|---|---|
| B₆S | 2545 | 11.44 | 7.40 | 8.12 | 1162 | 2.59 | 48.37 | 1.08 | 3851 |
| B₆Se | 3604 | 9.61 | 6.29 | 6.89 | 970 | 2.12 | 45.92 | 1.04 | 3565 |

*Calculated at 300 K

According to the Anderson formula [69], $\Theta_D$ is largely dependent on the $v_m$ which is linked to shear modulus ($G$) and bulk modulus ($B$). It is seen that the calculated value of $\Theta_D$ are found to be 1162 K and 970 K for the compositions of B₆S and B₆Se, respectively. The high value of $\Theta_D$ for the B₆S compared to B₆Se is due to large value of $v_m$, an indication of high minimum thermal conductivity, ($K_{min}$) as well as lattice thermal conductivity ($k_{ph}$). These values are much higher than those of many useful materials such as Sapphire (1047 K) and some other materials, which are widely used in various application sectors [70].

3.8 *Lattice thermal conductivity*

The lattice thermal conductivity ($k_{ph}$) is a measure of ability to conduct heat by lattice vibrations in a solid. The materials of high lattice thermal conductivity are widely used in heat sink applications, and materials with low lattice thermal conductivity are used as thermal insulation. The empirical formula derived by Slack [71] has been used to determine the $k_{ph}$ of B₆X (X = S, Se).

$$K_{ph} = A(\gamma)\frac{M_{av}\Theta_D^3\delta}{\gamma^2 n^{2/3} T}$$

$$\gamma = \frac{3(1+v)}{2(2-3v)}$$

where, $\gamma$ is the Grüneisen parameter associated with the anharmonicity of phonons. The value of $\gamma$ is estimated and found to be low for the materials of interest (see Table 8). The coefficient $A(\gamma)$ is calculated using the estimated Grüneisen parameter as follows.

$$A(\gamma) = \frac{4.85628 \times 10^7}{2(1 - \frac{0.514}{\gamma} + \frac{0.228}{\gamma^2})}$$

On account of modified Clarke's model [72], the theoretical lower limit of intrinsic thermal conductivity can be presented as:

$$K_{min} = k_B v_m \left(\frac{M}{n\rho N_A}\right)^{-\frac{2}{3}}$$

The minimum thermal conductivity, $K_{min}$ and lattice thermal conductivity, $k_{ph}$ at temperature 300 K have been calculated along with the Grüneisen parameter, $\gamma$ and are listed in Table 8. It is seen that the compound B$_6$S shows higher $k_{ph}$ (48.37 W/mK) than that (45.92 W/mK) of B$_6$Se as the bonding strength and $\Theta_D$ of this particular compound is also higher than those of B$_6$Se compound. Materials with high thermal conductivity are used as heat sinks in the laptop or any microelectronic device generating heat. Aluminum /aluminum alloys, copper, boron-aluminium, aluminium nitride, diamond, SiC and some others compounds with thermal conductivity value (> 145 W/mK) are very well known materials used for heat sink applications [73]. The values of $k_{ph}$ (45 - 48 W/mK) of our studied compounds are much lower than that of these materials. The thermal conductivity of B$_6$X (X = S, Se) are expected to be much higher at high temperatures since the Debye temperatures of the compounds under investigation are quite high compared to room temperature (300 K). However, it is interesting to note that the estimated value of $k_{ph}$ at 300 K are higher than that of well-known MAX phase compounds, bronze and comparable with that of red brass, and aluminum bronze [55,73,74].

The titled compounds are synthesized at 2700 K, thus, the melting temperature of these compounds should be high, at least greater than the synthesis temperature. To predict application at high temperature technology, melting temperature ($T_m$) should be known. The $T_m$ of semiconducting materials can be estimated using the relation [75,76]: $T_m = 412 + [8.2 \times (C_{11} + C_{12}^{1.25})]$. The estimated $T_m$ for B$_6$S and B$_6$Se are 3851 K and 3565 K, respectively, which are much higher than that of synthesis temperature.

3.9 *Heat capacities and thermal expansion coefficients*

The phonon specific or heat capacity at constant volume ($C_v$) of the compounds can be calculated using the quasi-harmonic Debye model [77–80]:

$$C_v = 9nN_A k_B \left(\frac{T}{\Theta_D}\right) \int_0^{x_D} dx \frac{x^4}{(e^x - 1)^2}$$

where, $x_D = \frac{\Theta_D}{T}$ ; and $n$, $N_A$ and $k_B$ are the number of atoms per formula unit, the Avogadro's number and the Boltzmann constant, respectively. The linear thermal expansion coefficient ($\alpha$) and specific heat at constant pressure ($C_p$) are calculated using following relations [77]:

$$\alpha = \frac{\gamma C_v}{3 B_T v_m} \text{ and } C_p = C_v(1 + \alpha \gamma T)$$

where, $B_T$, $v_m$ and $\gamma$ are the isothermal bulk modulus, molar volume and Grüneisen parameter, respectively.

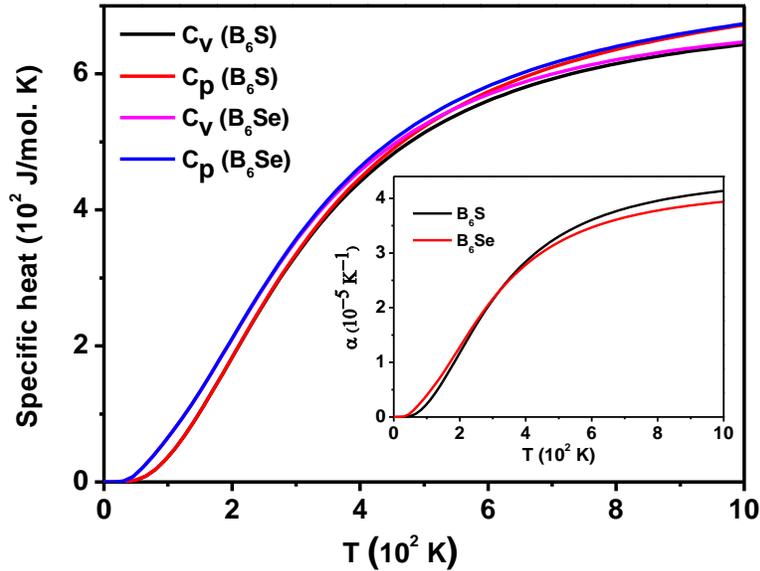

Fig. 12: Temperature dependence of specific heat $C_v$ at constant volume and $C_p$ at constant pressure for the compounds $B_6X$ (X = S, Se). The inset shows the temperature dependent linear thermal expansion coefficient, α, for the $B_6X$ (X = S, Se) compounds.

The temperature dependence of specific heats, $C_v$, $C_p$ and α for the $B_6X$ (X = S, Se) compounds are illustrated in Fig. 12. The parameters are evaluated in the temperature range of 0-1000 K where the quasi-harmonic Debye model is assumed to be valid and no structural phase transitions are expected for the compounds considered here. The phonon thermal softening happens with increasing temperature. As a result the heat capacity also increases with increasing temperature. The heat capacities increase rapidly at lower temperature regime and follow the Debye-$T^3$ power-law [81] and they approach to the Dulong-Petit (DP) ($3nN_Ak_B$) limit at high temperature regime where $C_v$ and $C_p$ do not depend strongly on temperature [32,82] any more.

The anharmonicity in the lattice dynamics can be measured by the thermal expansion coefficient (α) of materials which arises for the difference between the specific heats $C_p$ and $C_v$. The value of α for the compounds are estimated and shown in the inset of Fig. 12. It is seen that the α rises rapidly up to temperature 400 K and then gradually increases in the temperature ranging from 400 K to to 600 K and then appears to saturate. The values of α are found to be 2.20× $10^{-5}$ $K^{-1}$ and 2.18× $10^{-5}$ $K^{-1}$, for the $B_6S$ and $B_6Se$ compounds at temperature 300 K, respectively.

To use a material for heat sink application, high value of thermal conductivity, low thermal expansion coefficient and low density are prerequisites. Based on the present results of thermal properties, it can be concluded that comparatively high value of Debye temperature ($\Theta_D$), minimum thermal conductivity ($K_{min}$), lattice thermal conductivity ($k_{ph}$), and low thermal expansion coefficient and low density suggest that both the boron-rich chalcogenides might be used as heat sink material.

## 4. Conclusions

The state-of–the-art density functional theory has been employed to calculate the mechanical, electronic, optical, thermal and dynamical properties of novel boron-rich chalcogenides $B_6X$ (X = S, Se) for the first time. The band structure calculations have disclosed wide indirect band gap semiconducting features for both the compounds. The analysis of the Mulliken bond population and electron density difference reveals that strong covalent bond between B-S/Se atoms are prominently at work. The absence of negative phonon frequency and the fulfillment of mechanical stability criteria of orthorhombic structure confirmed that both the compounds are dynamically and mechanically stable. The high hardness and covalent bonding properties are supported by the PDOS profiles. The hardness values of $B_6Se$ ($H_{macro}$ = 33.02 GPa, $H_{micro}$ = 35.57 GPa) are higher than those of $B_6S$ ($H_{macro}$ = 31.02 GPa, $H_{micro}$ = 33.60 GPa) have been explained successfully by shifting of the peak position of total density of states in the low energy side from the Fermi level. The different anisotropic indices attest the anisotropic nature of the titled compounds. Different optical functions (dielectric constants, photoconductivity, absorption coefficient, reflectivity, refractive index and loss function) are calculated and their spectral behaviors are discussed discretely. The features of optical spectra make the compounds plausible candidate for the applications in medical autoclave and even in optoelectronic devices in the mid-UV region. Technologically important thermal parameters such as Debye temperature, thermal conductivity, heat capacities and thermal expansion coefficient are calculated and analyzed in details suggesting that both the boron-rich chalcogenides are highly thermally conductive.

We hope that the results presented herein will provide as a reference for future experimental and further theoretical studies on these fascinating boron-rich chalcogenides systems for the assessment of their suitability in various device applications.

**Acknowledgements**

Authors are grateful to the Department of Physics, Chittagong University of Engineering & Technology (CUET), Chattogram-4349, Bangladesh, for providing the computing facilities for this work.

**Data availability**

The datasets generated during the current study are available from the corresponding authors on a reasonable request.

**Conflict of Interest**

The authors declare that they have no known competing financial interests or personal relationships that could have appeared to influence the work reported in this paper.